# Analyzes of the Distributed System Load with Multifractal Input Data Flows

Kirichenko Lyudmyla[1], Radivilova Tamara[2]

1. EM Department, Kharkiv National University of Radioelectronics, UKRAINE, Kharkiv, 14 Nauki ave. email: lyudmyla.kirichenko@nure.ua

2. ICE Department, Kharkiv National University of Radioelectronics, UKRAINE, Kharkiv, 14 Nauki ave., email: tamara.radivilova@gmail.com

*Abstract*: **The paper proposes a solution an actual scientific problem related to load balancing and efficient utilization of resources of the distributed system. The proposed method is based on calculation of load CPU, memory, and bandwidth by flows of different classes of service for each server and the entire distributed system and taking into account multifractal properties of input data flows. Weighting factors were introduced that allow to determine the significance of the characteristics of server relative to each other. Thus, this method allows to calculate the imbalance of the all system servers and system utilization. The simulation of the proposed method for different multifractal parameters of input flows was conducted. The simulation showed that the characteristics of multifractal traffic have a appreciable effect on the system imbalance. The usage of proposed method allows to distribute requests across the servers thus that the deviation of the load servers from the average value was minimal, that allows to get a higher metrics of system performance and faster processing flows.**

## I. INTRODUCTION

Due to the massive spread of distributed computing system the problem of their effective use has become relevant. One aspect of this problem is the effective planning and allocation of tasks within a distributed computing system in order to optimize of resources utilization and reduce the computation time. Quite often there is a situation in which a portion of computational resources idle, while another portion of resources is overloaded and a large number of tasks awaiting execution in the queue.

To optimize resource utilization, reducing the time of service requests, horizontal scaling (dynamic addition/removal of devices) and failover (backup) the method of uniform distribution of tasks between multiple network devices (eg, servers) is applying and called Load Balancing [1-3].

When new tasks come the software that implements the balance must decide on what compute node should perform calculations related to this new task. In addition, balancing involves the transfer (migration - migration) of task's part from the most loaded compute nodes to less load nodes. In the performance of tasks processors exchange among themselves by communication messages. In the case of low communication costs, some processors (computers) may be idle while others are loaded. There will also be inappropriate high cost of communication. Consequently, balancing strategy should be such that the computing nodes have been loaded quite evenly, but also communication environment must not be overloaded.

The most famous studies in the field of balancing, theoretical research and development of fundamentals of load distribution, the creation of a mathematical apparatus, models and methods of management for load balancing in distributed systems are engaged by scientists like V. Cardellini [1, 2], E.I. Ignatenko [3], Hisao Kameda, Lie Li [4], H. Chen, F. Wang [5], S.Keshav [6], Xing-Guo Luo, Xing-Ming Zhang [7] and other researchers.

Numerous studies of processes in information networks have shown that network traffic has the property of scale invariance (self-similarity). Self-similar traffic has a special structure, conserved on many scales – there are large amount of bursts with a relatively small average level of traffic. These bursts cause significant delay and packet loss, even when the total load of all flows is far from the maximum permissible values.

Self-similar properties discovered in the local and global networks, particularly traffic Ethernet, ATM, applications TCP, IP, VoIP and video streams. The reason for this effect lies in the features of the distribution of files on servers, their sizes, the typical behavior of users [8-10].

It turns out that data streams originally did not exhibit self-similarity properties, having treatment on the hub server and active network elements, begin feeding pronounced signs of self-similarity. The presence of self-similarity property in the transferred customer information flows has a great influence on the performance of distributed systems. A particularly important role it plays for services, providing the transmission of multimedia traffic, and real-time traffic. Thus, the actual task is the development and analysis of load balancing algorithm that takes into account the self-similarity of the traffic and load of each node and the entire distributed system. Now the multifractal properties of traffic are intensively studied. Multifractal traffic is defined as an extension of self-similar traffic due to take account of properties of second and higher statistics.

The purpose of work is to simulate dynamic load balancing in a distributed system based on the monitoring server load at various parameters of multifractal input traffic.

## II. MONITORING OF THE SYSTEM STATE

The problem of the load balancing occurs for the following reasons [1,3,11]:
- Non-uniform structure of objectives, the various logical processes require different computing power;
- Cluster structure is also not uniform, i.e., different computing nodes have different capacities;
- Inter-node communication structure is not homogeneous, since link connecting nodes may have different bandwidth characteristics.

Monitoring the status of servers and free bandwidth can be accomplished in three ways [12-17]:
- after each incoming request;
- at fixed intervals determined by a static algorithm;
- in the non-fixed time intervals determined by a dynamic algorithm.

The information obtained by the first method is the largest volume, since measurements are taken after each incoming request. In the second method, the amount of information constantly, but it is necessary to determine the information reading range, that the amount of information has not been excessive and insufficient. In the third method, the amount of information depends on the frequency of control intervals which must adapt to the structure of the incoming traffic due to its self-similar structure.

Before describing the load balancing strategy, which includes a comprehensive measurement of the total system imbalance level (degree of uniformity of load distribution between servers), some of the concepts and definitions must be entered.

The term "part" means the portion of CPU resources allocated to the task. If the task stands out more part of process than other tasks, this task gets more CPU resources from the fair part scheduler. CPU parts are not equivalent percent of the CPU resources.

Parts can determine the importance of workloads relative to other workloads. When assigning processor parts to task the most important is not the number of parts allocated to the task. Much more important is to know how many parts allocated to task in compared to the other tasks. It should also be take into account that many of these tasks will compete with this task for the CPU.

CPU usage, memory, and the channel will be regarded as dimensionless quantities, whose values are normalized and are in the range [0, 1]. Is generally accepted that if the average value of permanent load exceeds 0.70, it is necessary to find out the reason for such behavior of the system in order to avoid problems in the future; if the average load of the system is close to one, then an urgent need to find the cause and fix it.

## III. SELF-SIMILAR AND MULTIFRACTAL TRAFFIC'S PROPERTIES

Stochastic process $X(t)$, $t \geq 0$, with continuous real-time variable is said to be self-similar of index $H$, $0 < H < 1$, if for any value $a > 0$ processes $X(at)$, and $a^{-H}X(at)$, have same finite-dimensional distributions:

$$Law\{X(at)\} = Law\{a^H X(t)\}. \qquad (1)$$

The notation $Law\{\cdot\}$ means finite distribution laws of the random process. Index $H$ is called Hurst exponent. It is a measure of self-similarity or a measure of long-range dependence of process. For values $0,5 < H < 1$ time series demonstrates persistent behaviour. In other words, if the time series increases (decreases) in a prior period of time, then this trend will be continued for the same time in future. The value $H = 0,5$ indicates the independence (the absence of any memory about the past) time series values. The interval $0 < H < 0,5$ corresponds to antipersistent time series: if a system demonstrates growth in a prior period of time, then it is likely to fall in the next period.

The moments of the self-similar random process can be expressed as $E\left[|X(t)|^q\right] = C(q) \cdot t^{qH}$ where the quantity $C(q) = E\left[|X(1)|^q\right]$.

In contrast to the self-similar processes (1) multifractal processes have more complex scaling behavior $Law\{X(at)\} = Law\{M(a) \cdot X(t)\}$ where $M(a)$ is random function that independent of $X(t)$. In case of self-similar process $M(a) = a^H$.

For multifractal processes the following relation holds $E\left[|X(t)|^q\right] = c(q) \cdot t^{qh(q)}$ where $c(q)$ is some deterministic function, $h(q)$ is generalized Hurst exponent, which is generally non-linear function. Value $h(q)$ at $q = 2$ is the same degree of self-similarity H. Generalized Hurst exponent of monofractal process does not depend on the parameter q: h(q)=H.

As a characteristic of heterogeneity multifractal data flow in the work was proposed to calculate range of generalized Hurst exponent $\Delta h = h(q_{min}) - h(q_{max})$. For monofractal processes generalized Hurst exponent is independent of parameter q, and is a straight line: h(q)=H, Δh=0. The greater heterogeneity of the process, ie. large number of bursts present in the traffic, the greater the range Δh.

## IV. INTEGRATED MEASUREMENT SYSTEM IMBALANCE

The method of complex measure the total value of the system imbalance and the average value of each server imbalance has been developed [18, 19]. The proposed load balancing method is based on the complex external and internal methods of monitoring. The usage of external monitoring system allows to periodically test the network to determine the most congested segments. The usage of internal monitoring of node status allows to get an objective information about load of node and assembly components.

1. Average CPU utilization $CPU_i^u(T)$ of each $u$ processors of $i$-th server defined as the averaged CPU utilization during an observed period $T$. For example, if the observation period is 1 min. and the CPU load is recorded every 10 seconds, i.e. $CPU_i^u$ is the mean value of the six recorded values $i$-th server.

Similarly, the average utilization of $r$ memory $RAM_i^r(T)$, network bandwidth of $k$ channel $Net_i^k(T)$ of server $i$ can be defined.

2. Since the measured CPU load, memory and channel usage by the flow of $qs$ class differs from the CPU load average $CPU_i^{qs}$, memory $RAM_i^{qs}$ and channel $Net_i^{qs}$ by the flow of the class $qs$ by lack of time operating system, and switching between tasks, the value $CPU_i^{qsv}$, $RAM_i^{qsv}$, $Net_i^{qsv}$ that indicate CPU, memory, and channel load by flows of $qs$ class can be entered, measured by the accounting system or the operating system monitor.

CPU utilization by flow $qs$ class is calculated as follows:

$$CPU_i^{qs} = CPU_i^u \times f_{CPU}^{qs}. \qquad (2)$$

where $f_{CPU}^{qs}$ - the part of total use $u$-th processor, which can be attributed to the $qs$ class. The parameter $f_{CPU}^{qs}$ is calculated as follows: $f_{CPU}^{qs} = CPU_i^{qsv} \Big/ \sum_{\forall qs} CPU_i^{qsv}$.

Similarly, values of memory load and bandwidth based on classes $qs$ flows are the calculated.

Memory load by flow $qs$ class is calculated as follows:

$$RAM_i^{qs} = RAM_i^r \times f_{RAM}^{qs}, \qquad (3)$$

where $f_{RAM}^{qs}$ - the part of total use $r$-th memory, which can be attributed to the $qs$ class. The parameter $f_{RAM}^{qs}$ is calculated as follows: $f_{RAM}^{qs} = RAM_i^{qsv} \Big/ \sum_{\forall qs} RAM_i^{qsv}$.

Bandwidth network by flow $qs$ class is calculated as follows:

$$Net_i^{qs} = Net_i^k \times f_{Net}^{qs}, \qquad (4)$$

where $f_{Net}^{qs}$ - the part of total use $k$-th channel, which can be attributed to the $qs$ class. The parameter $f_{Net}^{qs}$ is calculated as follows: $f_{Net}^{qs} = Net_i^{qsv} \Big/ \sum_{\forall qs} Net_i^{qsv}$.

3. Introduce average utilization of all servers CPU in a distributed system. Let $n_i$ is the amount of CPUs of $i$-th server, $CPU_i^{n_i}$ average load CPU of $i$-th server, then

$$CPU_u^{All} = \frac{\sum_i^N CPU_i^u CPU_i^{n_i}}{\sum_i^N CPU_i^{n_i}}. \qquad (5)$$

where N is the total number of servers in a system.

Similarly, the average utilization of memory $RAM_i^{m_i}$, network bandwidth $Net_i^{k_i}$ of $i$-th server, all memories $RAM_r^{All}$, and all network bandwidth $Net_k^{All}$ in a system can be defined.

$$RAM_r^{All} = \frac{\sum_i^N RAM_i^r RAM_i^{m_i}}{\sum_i^N RAM_i^{m_i}}. \qquad (6)$$

$$Net_k^{All} = \frac{\sum_i^N Net_i^k Net_i^{k_i}}{\sum_i^N Net_i^{k_i}}. \qquad (7)$$

4. The imbalance value of all CPUs. Using dispersion formulas, the imbalance value of all CPUs in distributed system is defined as

$$IMB_{CPU} = \sum_i^N (CPU_i^u - CPU_u^{All})^2 \qquad (8)$$

Similarly, imbalance values of memory $IMB_{RAM}$ and network bandwidth $IMB_{Net}$ can be calculated.

$$IMB_{RAM} = \sum_i^N (RAM_i^r - RAM_r^{All})^2, \qquad (9)$$

$$IMB_{Net} = \sum_i^N (Net_i^k - Net_k^{All})^2. \qquad (10)$$

5. Lets introduce complex value $IMB_i$ load imbalance $i$-th server, which takes into account all three server resource. Using the formula for calculating the variance as a measure of non-uniformity, the integrated value of load imbalance $i$-th server can be defined as:

$$IMB_i = a(CPU_i^u - CPU_u^{All})^2 + b(RAM_i^r - RAM_r^{All})^2 + \\ + c(Net_i^k - Net_k^{All})^2 \qquad (11)$$

Parameters $a, b, c$ represent weighting coefficients for the processor, memory and network bandwidth, respectively, which are selected by experimentation, so that $a + b + c = 1$ and depend on the tasks and the system structure. $IMB_i$ is used to indicate load imbalance level by comparing the coefficient of CPU utilization, memory and bandwidth. Value $IMB_i \to \min$ should be minimize.

Then the total value of the imbalance of all servers in the system is given as:

$$IMB_{tot} = \frac{1}{N} \sum_i^N IMB_i. \qquad (12)$$

6. Average duration of work with the same amount tasks allows to compare different scheduling algorithms.

7. The processing period on $i$-th server is defined as the maximum load on $i$-th server. The treatment period in the system is defined as the average load on all servers.

8. Efficiency is defined as the average load on any server.

Thus, the method of complex measurements of general level of integrated system imbalance has been developed for scheduling resources, as well as the average level of each server imbalance.

## V. THE RESULTS OF THE SIMULATION

For the simulation of load balancing of distributed system with an input multifractal flows the software was created, written in Python. This software allows the simulation of work load balancing system by using different balancing algorithms using the proposed dynamic load balancing method.

Multifractal traffic generated as described in [8] and input to the system. Requests coming from the external network, forming a multifractal traffic and sent to the load balancer, which in turn regulates the flow of tasks by using the selected balancing policy and give tasks to servers.

Also a system has Secondary Load Balancer, which provides fault tolerance, restoring balancer if it could not stand the load. The usage of this structure makes it possible to distribute the load between the servers by means of interaction between the components of the program with each other.

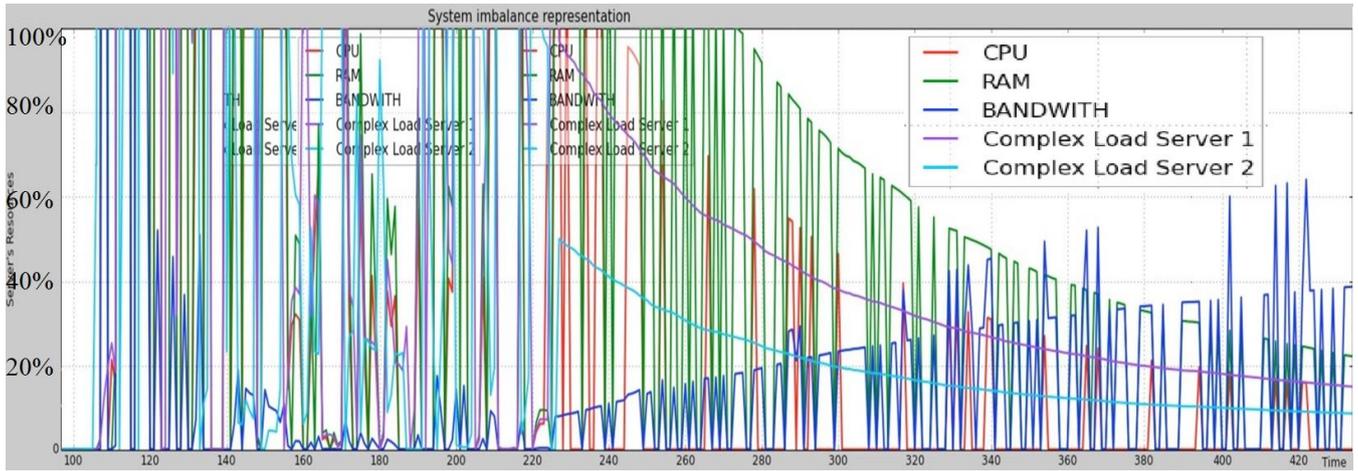

Fig.1. System imbalance for traffic parameters H=0.9 and Δh= 4

The proposed load balancing method is based on the complex internal and external monitoring methods. Using an external monitoring system allows to periodically test the network to determine most loaded segments. Using the internal monitoring status of node allows to get an objective data about node's load and information about individual assembly components load.

During experiments it was found that the data is changed slightly by increasing the number of cluster nodes and the number of servers in them. Therefore, to save time of generate traffic and calculations two clusters with the number of servers equal to six were chosen in each cluster. As usual clusters are heterogeneous and they have server with different capacity, the following servers parameters were selected in each cluster:

$\mu_{1n}(CPU_{1n}, RAM_{1n}, Net_{1n})$, $\mu_{2n}(CPU_{2n}, RAM_{2n}, Net_{2n})$, $1 \leq n \leq 6$, where $CPU_{1,2n} = 400$, $RAM_{1,2n} = 450$, $Net_{1,2n} = 300$, when $n = 1,3,5$. $CPU_{1,2n} = 300$, $RAM_{1,2n} = 350$, $Net_{1,2n} = 250$, when $n = 2,4$; $CPU_{16} = 500$, $RAM_{16} = 550$, $Net_{16} = 350$; $CPU_{26} = 600$, $RAM_{26} = 650$, $Net_{26} = 400$.

Parameters $a, b, c$ (11) that represent weighting coefficients for the processor, memory and network bandwidth, respectively, were selected equivalent. Figures 1 show the change of processors, memory and bandwidth imbalance and complex value load imbalance of each cluster when the multifractal parameters H=0.9 и Δh=4.

The table 1 shows the average values of system imbalance and time normalization of the system and its entrance into equilibrium state of depending on the change of multifractal characteristics of traffic.

Analyzes of simulation results shows that the first 230 seconds the system is in an unstable state, and resources are spent inefficiently, and after 230 seconds the imbalance of the system begins to decrease, the average utilization of resources are decreases too . After 400 seconds of work, the system comes to a stable state, and the value of imbalance remains practically unchanged.

Research shows that the imbalance of the system depends essentially on the multifractal traffic characteristics. For small values of H and small generalized Hurst exponent the balancing system reaches equilibrium and imbalance value to minimum. With increasing Hurst index over time the imbalance of the system is reduced by the same small values of heterogeneity and the system reaches equilibrium.

TABLE 1. CHANGE OF IMBALANCE DEPENDING ON THE MULTIFRACTAL PARAMETERS

| Multifractal parameters | Time of equilibrium, sec. | $IMB_{tot}$ |
|---|---|---|
| H=0,6, Δh=1,5 | 110 | 0.2 |
| H=0,6, Δh=2 | 130 | 0.2 |
| H=0,6, Δh=4 | 200 | 0.5 |
| H=0,6, Δh=6 | 310 | 0.58 |
| H=0,7, Δh=1,5 | 110 | 0.3 |
| H=0,7, Δh=2 | 140 | 0.33 |
| H=0,7, Δh=4 | 220 | 0.55 |
| H=0,7, Δh=6 | 320 | 0.65 |
| H=0,8, Δh=1,5 | 160 | 0.34 |
| H=0,8, Δh=2 | 180 | 0.4 |
| H=0,8, Δh=4 | 240 | 0.63 |
| H=0,8, Δh=6 | 400 | 0.7 |
| H=0,9, Δh=1,5 | 170 | 0.4 |
| H=0,9, Δh=2 | 180 | 0.45 |
| H=0,9, Δh=4 | 240 | 0.7 |
| H=0,9, Δh=6 | >500 | ≈1 |

However, with increasing heterogeneity of traffic even with the small Hurst exponent values system is not balanced, and the imbalance of the system has a large burstability. For the large Hurst exponent values and large heterogeneity the balancing system is in unstable state and the imbalance value tends to maximum, which means the maximum load of resources. It can be concluded that the self-similar properties of traffic have significant effect on the state of the system, but traffic heterogeneity has even more influence. The more heterogeneous traffic, the more resources is necessary for its processing.

## VI. Conclusion

In this work it is proposed a solution of actual scientific task of load nodes evaluation of the distributed system. As the evaluation of nodes resources the characteristics loading of the processor, memory and bandwidth are offered. In the proposed method, the average load of CPU, memory and bandwidth are calculated. It is based on the multifractal properties of traffic, which is measured by accounting system or operating system monitor. This method allows to calculate the loading of the processor, memory and bandwidth for multifractal flows of different classes of service for each server separately and for the entire distributed system.

The method allows to count the imbalance of all processors of distributed system, and memory, and bandwidth. Weighting factors are introduced that allow to determine the significance of the characteristics of server relative to each other. The complex value of server load imbalance taking into account the weighting factors for the processor, memory, and network bandwidth is also introduced. Thus, this method allows to calculate the imbalance of the all system servers and system utilization.

In further it is planned to carry out experiments on the work of proposed algorithm and standard balancing algorithm and compare them by the quantity of lost data, and the average waiting time of tasks in the system.

## References

[1] V. Cardellini, M. Colajanni, P. S. Yu. "Dynamic Load Balancing on Web-server Systems", IEEE Internet Computing, Vol.3, No.3, pp.28-39, 1999.

[2] V. Cardellini, "A performance study of distributed architectures for the quality of web-services", Proceedings of the 34th Conference on System Sciences, vol.10, pp.213-217, 2001.

[3] E.I. Ignatenko, V.I. Bessarab, I.V. Degtyarenko, "An adaptive algorithm for monitoring network traffic cluster in the load balancer", Naukovi pratsi DonNTU, Vol.21(183), pp. 95-102, 2011.

[4] Hisao Kameda, Lie Li, Chonggun Kim, Yongbing Zhang, "Optimal Load Balancing in Distributed Computer Systems", Springer, Verlag London Limited, 1997, P. 238.

[5] H. Chen, F. Wang, N. Helian, G. Akanmu, "User-priority guided min-min scheduling algorithm for load balancing in cloud computing", National Conference Parallel Computing Technologies (PARCOMPTECH), pp.1-8, 2013.

[6] S.Keshav, "An Engineering Approach to Computer Networking", Addison-Wesley, Reading, MA, pp. 215-217, 1997.

[7] Jing Liu, Xing-Guo Luo, Xing-Ming Zhang, Fan Zhang and Bai-Nan Li, "Job Scheduling Model for Cloud Computing Based on Multi-Objective Genetic Algorithm", IJCSI International Journal of Computer Science, v.10(1), № 3, pp.134-139, 2013.

[8] L. Kirichenko, T. Radivilova, E. Kayali, "Modeling telecommunications traffic using the stochastic multifractal cascade process", Problems of Computer Intellectualization ed. K. Markov, V. Velychko, O. Voloshin, Kiev–Sofia: ITHEA, pp. 55–63, 2012,.

[9] Zhenyu Na, Yi Liu, Yang Cui, Qing Guo, "Research on aggregation and propagation of self-similar traffic in satellite network", International Journal of Hybrid Information Technology, Vol.8, No.1, pp. 325-338, 2015.

[10] L. Kirichenko, I. Ivanisenko, T. Radivilova, "Investigation of Self-similar Properties of Additive Data Traffic", CSIT-2015 X-th International Scientific and Technical Conference «Computer science and information technologies», Lviv, Ukraine, pp. 169-172, 14 – 17 September 2015.

[11] Gregor Roth, "Server load balancing architectures, Part 1: Transport-level load balancing" [Online]. Available: http://www.javaworld.com/article/2077921/architecture-scalability/server-load- balancing-architectures- -part- 1--transport-level- load-balancing.html. [Accessed: 14- Jan-2017].

[12] Rudra Koteswaramma, "Client-Side Load Balancing and Resource Monitoring in Cloud", International Journal of Engineering Research and Applications (IJERA), Vol.2(6), pp. 167-171, 2012.

[13] Dhinesh Babu L.D., P. Venkata Krishna, Honey bee behavior inspired load balancing of tasks in cloud computing environments, Applied Soft Computing, Volume 13, Issue 5, pp. 2292–2303, 2013.

[14] Zhihao Shang, Wenbo Chen, Qiang Ma, Bin Wu, "Design and implementation of server cluster dynamic load balancing based on OpenFlow", Awareness Science and Technology and Ubi-Media Computing (iCAST-UMEDIA), pp. 691 – 697, 2013.

[15] Martin Randles, David Lamb, A. Taleb-Bendiab, "A Comparative Study into Distributed Load Balancing Algorithms for Cloud Computing", IEEE 24th International Conference on Advanced Information Networking and Applications Workshops, pp. 551-556, 2010.

[16] Thomas Erl, Robert Cope, Amin Naserpour, "Cloud Computing Design Patterns", Prentice Hall, Ed.1st., p.592, 2015.

[17] Steven Levine, John Ha, Red Hat Enterprise Linux 6. Load Balancer Administration. Load Balancer Add-on for Red Hat Enterprise Linux Edition 6, Red Hat, Inc., 2016. – 63 p.

[18] Wenhong Tian, Yong Zhao. Optimized Cloud Resource Management and Scheduling: Theories and Practices: - Morgan Kaufman, 2014. – 284 p.

[19] M. Ashraf Iqbal, Joel H. Saltz, Shahid H. Bokhari, Performance Tradeoffs in Static and Dynamic Load Balancing Strategies, NASA, March 1986. – p.28.